\documentclass[prc,twocolumn,nofootinbib,superscriptaddress,showpacs,aps]{revtex4-1}

\usepackage{amsmath}
\usepackage{amssymb}
\usepackage{mathtools}
\usepackage{graphicx}
\usepackage{dcolumn,dsfont}
\usepackage{bm,braket}
\usepackage{hyperref}
\usepackage{isotope}
\usepackage{xspace}
\usepackage{color}
\usepackage{multirow}
\usepackage{simplewick}
\usepackage[activate={true,nocompatibility},final,tracking=true,
kerning=true,spacing=true,factor=1100,stretch=10,shrink=10]{microtype}

\usepackage[all]{nowidow}
\hypersetup{
     colorlinks   = true,
     linkcolor = cyan,
     citecolor = cyan,
     menucolor = black,
     urlcolor = cyan  
}

\begin{document}
\renewcommand{\vec}[1]{\mathbf{#1}}
\newcommand{\clebschG}[6]{\mathcal{C}_{#1 #2 #3 #4}^{#5 #6}}  
\newcommand{\kF}{k_{\rm F}}
\newcommand{\MeV}{\,\mathrm{MeV}}
\newcommand{\GeVi}{\,\mathrm{GeV}^{-1}}
\newcommand{\fm}{\,\mathrm{fm}}
\newcommand{\fmi}{\,\mathrm{fm}^{-1}}
\newcommand{\fmiq}{\,\mathrm{fm}^{-3}}
\newcommand{\clebschGA}[6]{
  \left(
  \begin{array}{cc|c}
  #1 & #3 & #5 \\
  #2 & #4 & #6
  \end{array}
  \right)
}
\newcommand{\sixJsymbsixA}[6]{
	\begin{Bmatrix}
  #1 & #3 & #5 \\
  #2 & #4 & #6
	\end{Bmatrix}
}  
\newcommand{\nineJsymbA}[9]{
	\begin{Bmatrix}
  #1 & #4 & #7 \\
  #2 & #5 & #8\\
  #3 & #6 & #9 
	\end{Bmatrix}
}  

\newcommand{\oneSzero}{$\prescript{1\!}{}{S}_0$\xspace}
\newcommand{\onePone}{$\prescript{1\!}{}{P}_1$\xspace}
\newcommand{\threeSone}{$\prescript{3\!}{}{S}_1$\xspace}
\newcommand{\threePzero}{$\prescript{3\!}{}{P}_0$\xspace}
\newcommand{\threePone}{$\prescript{3\!}{}{P}_1$\xspace}
\newcommand{\threePtwo}{$\prescript{3\!}{}{P}_2$\xspace}
\newcommand{\threePFtwo}{$\prescript{3\!}{}{P}_2\!-\!\prescript{3\!}{}{F}_2$\xspace}
\newcommand{\threeFtwo}{$\prescript{3\!}{}{F}_2$\xspace}
\newcommand{\oneDtwo}{$\prescript{1\!}{}{D}_2$\xspace}
\newcommand{\oneGfour}{$\prescript{1\!}{}{G}_4$\xspace}

\title{Pairing in neutron matter: New uncertainty estimates and three-body forces}

\author{C.\ Drischler}
\email[Email:~]{christian.drischler@physik.tu-darmstadt.de}
\affiliation{Institut f\"ur Kernphysik, Technische Universit\"at Darmstadt, 64289 Darmstadt, Germany}
\affiliation{ExtreMe Matter Institute EMMI, GSI Helmholtzzentrum f\"ur Schwerionenforschung GmbH, 64291 Darmstadt, Germany}

\author{T.\ Kr{\"u}ger}
\email[Email:~]{thomas.krueger@physik.tu-darmstadt.de}
\affiliation{Institut f\"ur Kernphysik, Technische Universit\"at Darmstadt, 64289 Darmstadt, Germany}
\affiliation{ExtreMe Matter Institute EMMI, GSI Helmholtzzentrum f\"ur Schwerionenforschung GmbH, 64291 Darmstadt, Germany}
 
\author{K.\ Hebeler}
\email[Email:~]{kai.hebeler@physik.tu-darmstadt.de}
\affiliation{Institut f\"ur Kernphysik, Technische Universit\"at Darmstadt, 64289 Darmstadt, Germany}
\affiliation{ExtreMe Matter Institute EMMI, GSI Helmholtzzentrum f\"ur Schwerionenforschung GmbH, 64291 Darmstadt, Germany}
 
\author{A.\ Schwenk}
\email[Email:~]{schwenk@physik.tu-darmstadt.de}
\affiliation{Institut f\"ur Kernphysik, Technische Universit\"at Darmstadt, 64289 Darmstadt, Germany}
\affiliation{ExtreMe Matter Institute EMMI, GSI Helmholtzzentrum f\"ur Schwerionenforschung GmbH, 64291 Darmstadt, Germany}
\affiliation{Max-Planck-Institut f\"ur Kernphysik, Saupfercheckweg 1, 69117 Heidelberg, Germany}

\begin{abstract}

We present solutions of the BCS gap equation in the channels \oneSzero
and \threePFtwo in neutron matter based on nuclear interactions
derived within chiral effective field theory (EFT). Our studies are
based on a representative set of nonlocal nucleon-nucleon (NN) plus
three-nucleon (3N) interactions up to next-to-next-to-next-to-leading
order (N$^3$LO) as well as local and semilocal chiral NN interactions
up to N$^2$LO and N$^4$LO, respectively. In particular, we investigate
for the first time the impact of subleading 3N forces at N$^3$LO on
pairing gaps and also derive uncertainty estimates by taking into
account results for pairing gaps at different orders in the chiral
expansion. Finally, we discuss different methods for obtaining
self-consistent solutions of the gap equation. Besides the widely-used
quasi-linear method by Khodel {\it et al.}~we demonstrate that the
modified Broyden method is well applicable and exhibits a robust
convergence behavior. In contrast to Khodel's method it is based on a
direct iteration of the gap equation without imposing an auxiliary
potential and is straightforward to implement.

\end{abstract}

\pacs{21.30.Fe, 21.65.Cd, 26.60.-c}

\maketitle

\section{Introduction}
\label{sec:intro}

A quantitative understanding of nuclear superfluidity is central for a
wide range of phenomena in nuclear systems, from the structure of
nuclei~\cite{Brin05Pairing,Dean03Pairing} to the cooling of neutron
stars~\cite{Yako04nscool,Page04mincool,Page11coolCasA}. In the inner
crust of neutron stars, neutron-rich nuclei form a crystal lattice
surrounded by a background liquid of neutrons in a superfluid state
(see, e.g., Ref.~\cite{Geze14pairing} for a review on superfluidity in
neutron stars). At densities up to $0.5 n_0$, with saturation density
$n_0 = 0.16\fmiq$, neutrons form Cooper pairs in the \oneSzero channel
since this channel provides the largest attractive interaction at low
momenta. Deeper inside the neutron star, in the outer core, the
density increases and at Fermi momenta of $\kF \sim 1.5 \fmi$ the
\oneSzero interaction becomes repulsive and the pairing gap closes in
this channel. At these densities the dominant attraction is in the
spin-triplet $P$-wave with total angular momentum $J = 2$, which is
coupled to the \threeFtwo channel. Beyond this density it is not
obvious to what extent present NN interactions are well constrained by
scattering data. Such uncertainties of the interaction are reflected
in results for the paring gaps.

Chiral EFT provides a systematic expansion for nuclear
forces~\cite{Epel09RMP,Mach11PR}, connecting the symmetries of quantum
chromodynamics to the interactions between nucleons. Recently, there
have been efforts to derive also local chiral
interactions~\cite{Geze13QMCchi,Geze14long} as well as semilocal
interactions using local regulators for long-range pion exchanges
while regulating the short-range contact interactions nonlocally in
momentum space~\cite{Epel15improved,Epel15NNn4lo}. These efforts
resulted in sets of NN interactions at different orders in the chiral
expansion for a given regulator, which enable more systematic
estimates of theoretical uncertainties due to the input nuclear forces.

Neutron pairing gaps in uniform matter have been investigated in the
BCS approximation based on chiral interactions, e.g., in
Refs.~\cite{Bald98TripPair,Hebe07Pairing,Hebe10nmatt,Dong13PairPNM,
Maur14pairing, Ding16scgfpair,Srin16PairPNM}. The BCS approximation
is particularly useful to test the sensitivity to nuclear
forces. However, we emphasize that there are important contributions
beyond the BCS approximation due to screening and vertex corrections,
which lead to significant changes to the BCS gaps (for a discussion
and further references see Ref.~\cite{Geze14pairing}). These are not
the focus of the present work and are not included in the
uncertainties studied here.

In the present paper, we study the zero-temperature pairing gap in
neutron matter in the \oneSzero and \threePFtwo channel based on new
local and semilocal NN interactions derived within chiral EFT up to
N$^2$LO and N$^4$LO, respectively. We also employ an improved method
for estimating uncertainties due to the truncation in nuclear
forces~\cite{Epel15improved,Epel15NNn4lo}, which is not based on
parameter variation but on an order-by-order analysis in the chiral
expansion. For the solution of the gap equation, we show that the
modified version of Broyden's method for solving general nonlinear
equations developed in Ref.~\cite{John88ModBroyn} is a powerful
method. In combination with the usual method of
Khodel~\textit{et~al.}~\cite{Khod01PairPNM} it allows to assess
systematically the iterative convergence. Furthermore, we study the
impact of 3N forces on the pairing gap at the level of normal-ordered
two-body contributions. Taking advantage of recent
developments~\cite{Hebe15N3LOpw,Dris15asym} we consider for the first
time N$^3$LO 3N contributions to the pairing interaction.

This paper is organized as follows. In Sec.~\ref{sec:calc_details} we
discuss details of our calculation, in particular the two independent
methods for solving the nonlinear gap equation and the treatment of 3N
forces. In Sec.~\ref{sec:results} we present our results for the
pairing gap in neutron matter in the \oneSzero and \threePFtwo
channel. We show results for the pairing gap using a free and a
Hartree-Fock (HF) single-particle spectrum and also for the effective
neutron mass as a function of density for all interactions
used. Finally, we summarize and give an outlook in
Sec.~\ref{sec:summary}.

\section{Calculational details}
\label{sec:calc_details}

\subsection{BCS gap equation}
\label{subsec:gap_eq}

The pairing gap is a $2\times 2$ matrix in single-particle spin space
obeying the BCS gap equation at
zero-temperature~\citep{Bald95DeutBCS}
\begin{equation} \label{eq:gap_start}
\Delta_{\alpha\alpha'}(\vec{k}) = 
-\sum \limits_{\substack{\beta,\beta' \\ \vec{k}'}}  \frac{\braket{\vec{k}\alpha\alpha'|V|\vec{k}' \beta\beta'} \Delta_{\beta\beta'}(\vec{k}')}{2 \sqrt{\xi^2(k')+\frac{1}{2} \text{Tr} \left[ \Delta \Delta^\dagger\right](\vec{k}') }} \, .
\end{equation}
The greek indices indicate the single-particle spin
states~$\ket{\pm}$, $\text{Tr}$ the trace in spin space and
$\xi(k)=\varepsilon(k)-\mu$ labels the single-particle energy, e.g.,
for a free spectrum $\varepsilon(k)=k^2/(2m)$ with the neutron mass
$m$, relative to the chemical potential $\mu$. Practically,
Eq.~\eqref{eq:gap_start} is solved in a partial-wave
representation. We review the decomposition in Appendix~\ref{app:PW}
in order to clarify the conventions and approximations used. As shown
in the appendix the angular integration can be carried out
analytically if the pairing gap in the energy denominator in
Eq.~\eqref{eq:gap_start} is averaged over all spacial directions:
\begin{equation} \label{eq:angle_av}
\begin{split}
\Delta^2(k) \equiv \frac{1}{2}\text{Tr} \left[ \Delta \Delta^\dagger \right] \xrightarrow{\text{av.}} &\frac{1}{2} \int \frac{d\Omega_\vec{k}}{4\pi} \text{Tr} \left[ \Delta \Delta^\dagger \right] \\
&= \sum \limits_{l,S,J} |\Delta_{lS}^{J}(k)|^2\, .
\end{split}
\end{equation}
In this approximation the partial-wave decomposed gap equation takes the form~\cite{Khod01PairPNM}
\begin{equation}\label{eq:gap_pw}
\Delta_{lS}^{J}(k) = - \int_{0}^{\infty}  \frac{dk' \, k'^2}{\pi}  \sum \limits_{l'}  \frac{ i^{l'-l} V_{ll'S}^{J}(k,k') \Delta_{l'S}^{J}(k')}{\sqrt{\xi^2(k')+  \sum \limits_{\tilde{l},\tilde{S},\tilde{J}} |\Delta_{\tilde{l}\tilde{S}}^{\tilde{J}}(k')|^2 }} \, .
\end{equation}
The different angular momenta $l,l' = |J \pm 1|$ are coupled in the
spin-triplet channel, whereas in the singlet channel we obtain $l' =
l$.  We note that due to the energy denominator the solutions of
$\Delta_{lS}^{J}$ are generally coupled, even if the interaction does
not couple these channels.  However, in practice Eq.~\eqref{eq:gap_pw}
can be solved to a very good approximation independently for fixed
quantum numbers $S$ and $J$, because they are dominated by the channel
in which the pairing interaction is most attractive at a given
density. This and angle-averaged gaps are commonly-used approximations
(note that the angle-averaging approximation is exact for the
\oneSzero channel).

In this paper, we solve Eq.~\eqref{eq:gap_pw} in pure neutron matter
for the most attractive channels of the nuclear interactions, the
spin-singlet channel \oneSzero and the triplet channel
\threePFtwo. The other channels in the triplet $P$-wave, \threePzero
and \threePone as well as in higher partial waves are less attractive
or even repulsive at the densities considered in this work. We have
checked that this also holds with the inclusion of 3N forces.
Following Eq.~\eqref{eq:angle_av} we plot the total gap $\Delta(k_{\rm
F}) = \sqrt{\sum_l \Delta_l^2(k_{\rm F})}$ evaluated on the Fermi
surface to estimate the pairing energy.

\subsection{Solving the gap equation}
\label{subsec:Gap_Eq}

The nonlinear gap equation~\eqref{eq:gap_pw} can be solved iteratively
until a self-consistent solution is obtained. However, such approaches
are computationally challenging and require more advanced
algorithms. The simplest and straightforward method that takes
directly the right-hand side of Eq.~\eqref{eq:gap_pw} $I[\ldots]$ in
the m-th iteration step,
\begin{subequations} \label{eq:direct_iter}
\begin{align}
{\vec{\Delta}_\text{out}^{(m)}} &= I\left[ {\vec{\Delta}_\text{in}^{(m)}} \right] \quad \text{with} \label{eq:apply_I}\\
{\vec{\Delta}_\text{in}^{(m+1)}} &= {\vec{\Delta}_\text{out}^{(m)}} \, ,  \label{eq:simple_update}
\end{align}
\end{subequations}
converges poorly, if at all. Instead, it typically converges to the
(mathematically also valid) trivial solution $\Delta = 0$, especially
if the nontrivial solution is small. We refer also to
Ref.~\cite{Bara08Broyden} for a general discussion of iterative
methods in the context of nuclear physics. In
Eqs.~\eqref{eq:direct_iter} we define a gap vector $\vec{\Delta}$
having as components the partial-wave $\Delta_l$ sampled each on a
Gauss momentum mesh with $N_p$ points. The basis size of this vector
is $N_p$ (spin singlet) and $2 \, N_p$ (spin triplet), respectively.

In addition to methodical convergence issues, also the evaluation of the
integral in Eq.~\eqref{eq:gap_pw} requires some care. Since the pairing gap is
typically a small energy scale, the integrand exhibits a strong peak structure
for momenta close to the Fermi surface. This quasi-singularity of the BCS gap
equation has to be treated carefully when evaluating the integral
numerically. We observe that Gauss quadrature converges only if multiple dense
integration meshes concentrated around the peak position are well distributed
over the entire interval (see also Ref.~\cite{Bald98TripPair}). The presence
of the peak makes the integral nevertheless quite sensitive to variations in
$\Delta(\kF)$ and can complicate obtaining a stable self-consistent solution.
In order to address these convergence issues, various methods have been applied in the
literature, for example the quasi-linear and linear methods of
Khodel {\it et al.}~\cite{Khod01PairPNM} and Krotscheck~\cite{Krot72SuFluNM},
or the instability analysis based on in-medium Weinberg
eigenvalues~\cite{Rama07WEVPair,Srin16PairPNM}. 

In order to assess the methodical convergence of our results we employ
two independent algorithms. These are discussed in detail in the next
sections. As it is often referred to, we quantify briefly the term
\emph{convergence}.  Let's consider a general solver that returns the
vectors $\vec{\Delta}_\text{in}^{(m)}$ and
$\vec{\Delta}_\text{out}^{(m)}$ after the $m$-th iteration, specified
by an update rule, for instance of the simple form
Eq.~\eqref{eq:simple_update}. The solver is stable if the norm of the
difference,
\begin{equation} \label{eq:F_diff}
{\vec{F}^{(m)}} = {\vec{\Delta}_\text{out}^{(m)}} - {\vec{\Delta}_\text{in}^{(m)}} \, ,
\end{equation}
decreases with $m$, eventually becomes smaller than an arbitrary fixed
threshold value and finally a self-consistent solution is found if
$|{\vec{F}^{(m)}}| = 0$. In practice, a small but finite threshold
serves as a break condition for the self-consistency cycle.  We check
the break condition for 5 to 10 additional iterations once it is
fulfilled.

\subsubsection{Khodel's method}

The method of Khodel {\it et al.} has been first presented in
Refs.~\cite{Khod96SolGap,Khod01PairPNM} and has since then been widely
used in nuclear physics (see, e.g., Refs.~\cite{Maur14pairing,
Srin16PairPNM} for recent applications). It is based on a
reformulation of the gap equation~\eqref{eq:gap_pw} such that the peak
of the integrand, causing the large sensitivity to $\Delta(\kF)$, is
removed. This is achieved by rewriting the potential $V_{ll'}(k,k')$
in a separable part
\begin{equation}
\phi_{ll'}(k) = \frac{V_{ll'}(k,\kF)}{v_{ll'}} \quad \text{and} \quad
\phi^T_{ll'}(k') = \frac{V_{ll'}(\kF,k')}{v_{ll'}} \, ,
\end{equation}
where the definition $v_{ll'} = V_{ll'}(\kF,\kF)\neq 0$ normalizes
$\phi_{ll'}(\kF) = \phi^T_{ll'}(\kF) = 1$, and a remainder
\begin{equation}\label{eq:remainder}
W_{ll'}(k,k') = V_{ll'}(k,k') - v_{ll'}\phi_{ll'}(k)\phi_{ll'}^T(k') \, ,
\end{equation}
which vanishes when at least one argument is on the Fermi
surface. This property is key to removing the peak. Inserting the
remainder~\eqref{eq:remainder} in the gap equation~\eqref{eq:gap_pw}
gives
\begin{align}\label{eq:Khodel1}
\Delta_l(k) &+ \sum \limits_{l'} i^{l'-l} \int \frac{dk' k'^2}{\pi} W_{ll'}(k,k') \frac{\Delta_{l'}(k')}{\sqrt{\xi^2(k')+\Delta^2(k')}} \nonumber\\
&= \sum \limits_{l'}D_{ll'}\phi_{ll'}(k) \, ,
\end{align}
with the coefficients defined as
\begin{equation}\label{eq:D_coeff}
D_{ll'} = - i^{l'-l} v_{ll'} \int \frac{dk k^2}{\pi} \frac{ \phi^T_{ll'}(k)\Delta_{l'}(k)}{\sqrt{\xi^2(k)+\Delta^2(k)}} \, .
\end{equation}
The partial-wave gap $\Delta_l$ in Eq.~\eqref{eq:Khodel1} can be written as linear combinations of shape functions $\chi_l^{l_1l_2}(k)$
\begin{equation}\label{eq:Gap_lin_comb}
\Delta_l(k) = \sum \limits_{l_1,l_2} D_{l_1l_2} \chi_l^{l_1l_2}(k) \, ,
\end{equation}
and thus one obtains an equation for the momentum dependence of the partial-wave gaps
\begin{align} \label{eq:khodel_chi}
\chi_l^{l_1 l_2}(k) &+ \sum_{l'} i^{l'-l} \int \frac{dk' k'^2}{\pi}
W_{ll'} (k,k')\frac{ \chi^{l_1l_2}_{l'}(k') }{\sqrt{\xi^2(k')+\Delta^2(k')}} \nonumber\\
&= \delta_{ll_1}\phi_{l_1l_2}(k) \, .
\end{align}
Since $W_{ll'}$ vanishes by construction if at least one argument is
on the Fermi surface, the integral in Eq.~\eqref{eq:khodel_chi} is
dominated by a momentum region where $\Delta(k)$ is far less important
than $\xi(k)$. The shape functions therefore only depend weakly on
$\Delta(k)$. This allows to treat Eq.~\eqref{eq:khodel_chi} to a good
approximation as quasi-linear, that means by approximating $\Delta(k)$
by a constant. Consequently, the momentum dependence of the gap
converges rapidly in Khodel's method and almost independently of their
magnitudes~\eqref{eq:D_coeff} due to the
separation~\eqref{eq:Gap_lin_comb}.

In practice, the iteration scheme works as
follows~\cite{Khod01PairPNM}: each momentum dependence is sampled on a
suitable Gauss mesh to ensure convergence of the quadrature. Given
$\Delta(k)$ from the previous iteration, one solves
Eq.~\eqref{eq:khodel_chi} for the shape functions $\chi_l^{l_1l_2}(k)$
by matrix inversion. For the first iteration a small constant value,
e.g., $\Delta(k)=1$~keV, serves as a suitable starting point. We
checked that our final results are independent of that choice.  The
coefficients $D_{ll'}$ can then be determined via
Eq.~\eqref{eq:D_coeff} combined with Eq.~\eqref{eq:Gap_lin_comb} using
a nonlinear solver such as the Newton-Raphson method. With the new
$D_{ll'}$ and $\chi_l^{l_1l_2}(k)$ Eq.~\eqref{eq:Gap_lin_comb} updates
the partial-wave gaps $\Delta_l(k)$. It follows directly from
Eq.~\eqref{eq:khodel_chi} that $\chi^{l_1l_2}_l(\kF) = \delta_{ll_1}$
for all $l_2$, so the total gap on the Fermi surface for the next
iteration step is simply $\Delta_l(\kF) = \sum_{l_2} D_{ll_2}$.  The
procedure is repeated until self-consistency is reached, typically
within a few iterations.

\subsubsection{Modified direct-iteration method}

As an alternative to Khodel's method, we solve for the gap by a
modified version of the direct-iteration method in
Eqs.~\eqref{eq:direct_iter}. Since Eq.~\eqref{eq:simple_update} is
known to be too simplistic, more advanced update rules are crucial to
achieve convergence. As a first step, the stability of the convergence
can be significantly improved by dampening the update
prescription. The simplest modification involves a linear
superposition of the input and output vector of the current iteration:
\begin{align}
{\vec{\Delta}_\text{in}^{(m+1)}} &= \alpha {\vec{\Delta}_\text{out}^{(m)}} + (1-\alpha) {\vec{\Delta}_\text{in}^{(m)}} \nonumber \\
&= {\vec{\Delta}_\text{in}^{(m)}} + \alpha {\vec{F}^{(m)}} \, ,
\end{align}
where $\alpha$ is the damping factor. We attempted to find suitable
values for $\alpha$ that lead to reliable convergence patterns for
various NN interactions over a typical range of densities. However, we
found that using simple mixing still results in too many
discontinuities of the gap as a function of density in order to be
useful in practice. These numerical artifacts had to be removed by
fine-tuning the damping factor for different densities. Hence,
reliable calculations for the gap require more sophisticated updates.

\begin{figure*}[t]
\includegraphics[page=1,scale=1.0,clip]{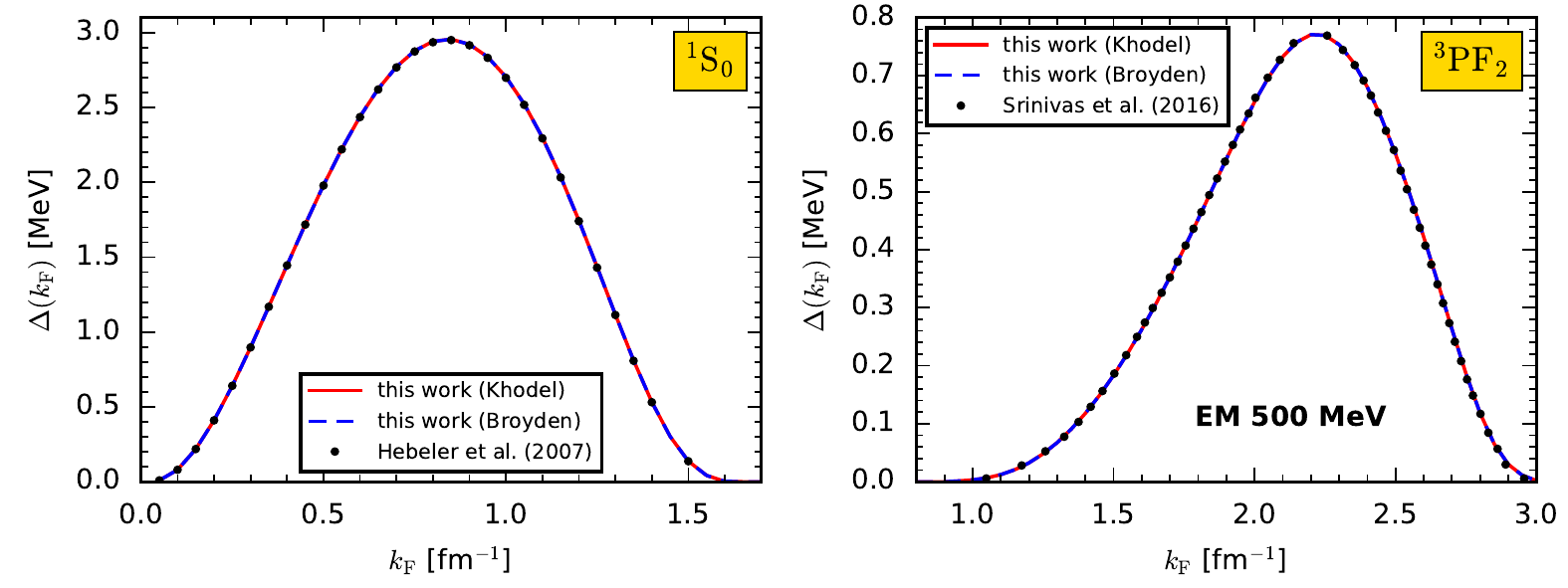}
\caption{\label{fig:gaps_benchmark}(Color online)
Comparison of the gap $\Delta$ as a function of Fermi momentum $k_{\rm F}$
in the \oneSzero~(left) and \threePFtwo~(right panel) channel obtained
using Khodel's method (red-solid) and via the new modified
direct-iteration method (blue-dashed line). Results are obtained with
the N$^3$LO NN potential
EM~500~MeV~\cite{Ente03EMN3LO,Mach11PR}. Values from
Refs.~\cite{Hebe07Pairing,Srin16PairPNM} are depicted as black
dots. The results are in very good agreement, especially, the two
methods of this work demonstrate excellent methodical convergence.}
\end{figure*}

We now demonstrate that Broyden's method for solving general nonlinear
equations is in particular well suited for the gap
equation~\eqref{eq:gap_pw}.  Specifically, we make use of a modified
version of Broyden's method developed in
Ref.~\cite{John88ModBroyn}. It is a fast, stable and computationally
efficient quasi-Newton-Raphson method with the advantage of a simple
but powerful update rule. The inverse of the Jacobian is approximated
by the knowledge of previous iterations without needing to store or to
process high-rank matrices. We review here briefly the ingredients to
obtain stable results for the gap and refer to the original
Ref.~\cite{John88ModBroyn} as well as to Ref.~\cite{Bara08Broyden} for
first applications to the nuclear many-body problem.

In the modified version of Broyden's method, the gap vector after the
$m$-th iteration is updated according to the rule
\begin{equation} \label{eq:Broy_update}
{\vec{\Delta}_\text{in}^{(m+1)}} = {\vec{\Delta}_\text{in}^{(m)}} + \alpha {\vec{F}^{(m)}} - \sum \limits_{n=1}^{m-1} w_n \gamma_{mn} {\vec{u}^{(n)}} \, ,
\end{equation}
with the definitions
\begin{align}
\gamma_{mn} &= \sum \limits_{k=1}^{m-1} c_{km} \beta_{kn} \, ,\\
\beta_{kn} &= (w_0^2 \mathds{1} + a )_{kn}^{-1} \, ,\\
c_{km} &= w_k \, {\delta \vec{F}^{(k)\dagger} \vec{F}^{(m)}} \, , \\
a_{kn} &= w_k w_n \, {\delta \vec{F}^{(n)\dagger} \delta \vec{F}^{(k)}} \, ,
\intertext{and}
{\vec{u}^{(n)}} &= \alpha \, {\delta \vec{F}^{(n)} }+  {\delta \vec{\Delta}^{(n)}} \, ,\\ 
{\delta \vec{\Delta}^{(n)}} &= \frac{{\vec{\Delta}_\text{in}^{(n+1)}}-{\vec{\Delta}_\text{in}^{(n)}}}{\left|{\vec{F}^{(n+1)}}-{\vec{F}^{(n)}}\right|} \, ,\\
{\delta \vec{F}^{(n)}} &= \frac{{\vec{F}^{(n+1)}}-{\vec{F}^{(n)}}}{\left|{\vec{F}^{(n+1)}}-{\vec{F}^{(n)}}\right|} \, ,
\end{align}
where $\delta \vec{F}^{(n)}$ is normalized, ${\delta
  \vec{F}^{(n)\dagger}\delta \vec{F}^{(n)}} = 1$. The procedure
requires to store ${\vec{\Delta}_\text{in}^{(m)}}$ and $\vec{F}^{(m)}$
of the current iteration as well as ${\vec{u}^{(m)}}$ and ${\delta
  \vec{F}^{(m)}}$ of all previous steps.  Since $a_{kn}$ is typically
of rank much smaller than that of the full Jacobian it can be stored
for efficiency.  Although the update rule~\eqref{eq:Broy_update}
includes simple mixing, the additional correction allows usually
larger damping factors $\alpha$, which typically leads to accelerated
convergence. Besides guesses for $\vec{\Delta}_\text{in}^{(1)}$ and
$\alpha$, the weights $w_m$ have to be chosen as well, whereas
$w_0=0.01$ needs to be sufficiently small~\cite{John88ModBroyn}. We
use $w_m=1$, $m \geqslant 1$, similar to Ref.~\cite{Bara08Broyden}. In
addition, Ref.~\cite{John88ModBroyn} suggested $w_m = \min \left( 1,
\sqrt{ (\vec{F}^{(m)\dagger}\vec{F}^{(m)} )^{-1}} \right)$ to promote
solutions of advanced convergence.

We show in Fig.~\ref{fig:gaps_benchmark} an exemplary benchmark for
the gap $\Delta(\kF)$ obtained with Khodel's~(red-solid) and with the
modified direct-iteration method~(blue-dashed lines) in comparison to
the literature (points)~\cite{Hebe07Pairing,Srin16PairPNM}. The gaps
are based on the N$^3$LO NN potential
EM~500~MeV~\cite{Ente03EMN3LO,Mach11PR} in the channels \oneSzero
(left) and \threePFtwo (right panel). We observe in general almost
perfect agreement (deviations are of order of $10$~eV) of the two
methods for the singlet as well as the triplet channel. We used the
same optimized Gauss mesh for the two methods. Furthermore, the
results in Fig.~\ref{fig:gaps_benchmark} agree well with the
literature, also in the regions of small gaps.  In practice, Khodel's
method requires typically $2$ to $3$ times fewer steps to converge
while the computational runtime is shorter for the modified
direct-iteration method due to its simplicity.  In rare cases the
modified direct-iteration method leads to apparent discontinuities in
the gap as a function of $\kF$. In all of our calculations we could
easily recover these by modifying slightly the damping factor
$\alpha$.  On the other hand, Khodel's method in its usual
implementation\footnote{Note that there is a modified version of
Khodel's method in Ref.~\cite{Khod96SolGap} accounting for
$V_{ll'}(k,k)=0$.} is naturally unstable if the $V_{ll'}(k,k)$ gets
small or has in particular nodes.

Based on these benchmarks, we conclude that the two algorithms are
equally reliable. Comparing the results of Khodel's method and the
modified direct-iteration method allows us to assess the methodical
convergence of our calculations. The results of the following
sections could therefore be obtained with either of the methods.

\subsection{Three-nucleon forces, normal ordering and single-particle energies}
\label{subsec:NO_3N_SPE}

Our calculations are based on NN and 3N interactions up to N$^3$LO in
the chiral expansion. The contributions from 3N forces are taken into
account at the normal-ordered two-body level. Normal-ordering with
respect to a given reference state allows to include the dominant 3N
contributions in terms of density-dependent two-body interactions
$\overline{V}_\text{3N}^\text{as}$~\cite{Holt10ddnn, Hebe10nmatt,
Carb14SCGFdd, Dris15asym, Well16DivAsym}.  Specifically, normal
ordering of 3N forces in neutron matter involves the summation of one
particle over occupied states in the Fermi sea:
\begin{equation} \label{eq:Veff_formal}
\overline{V}_\text{3N}^\text{as} = \text{Tr}_{\sigma_3} \int \frac{d\vec{k}_3}{(2\pi)^3} \mathcal{A}_{123} V_\text{3N} \, n_{\vec{k}_3} \bigg|_{\text{nnn}} \, ,
\end{equation}
where the Fermi-Dirac distribution function is given at zero
temperature by a simple step function, $n_\vec{k} = \theta \left( \kF
- |\vec{k}| \right)$, and the Fermi momentum $\kF$ associated with the
particle density by \mbox{$n=\kF^3/(3\pi^2)$}. The antisymmetrized 3N
interactions $\mathcal{A}_{123}V_\text{3N}$ used in this work are
regularized by the nonlocal regulator $f_\text{R}(p,q)=\exp [-((p^2+
3q^2/4) / \Lambda_{\text{3N}}^2)^{4}]$, where $p,q$ are the Jacobi
momenta and $\Lambda_{\text{3N}}$ is the 3N cutoff scale.

The contributions of 3N forces at N$^2$LO to the BCS pairing gap have
already been studied via normal ordering, see, e.g.,
Refs.~\cite{Hebe10nmatt,Dong13PairPNM,Maur14pairing,Srin16PairPNM}. The
calculation of $\overline{V}_\text{3N}^\text{as}$ can be performed
directly based on the operator structure of the 3N interactions as in
Refs.~\cite{Hebe10nmatt,Holt10ddnn}. However, this approach becomes
rather cumbersome for subleading 3N forces at N$^3$LO due to the
complex operator structure of 3N interactions at this order. In order
to study N$^3$LO 3N contributions we make use of recent
developments~\cite{Hebe15N3LOpw,Dris15asym} and evaluate the effective
NN potentials~\eqref{eq:Veff_formal} using the partial-wave
decomposition of the 3N forces. The partial-wave matrix elements of
the 3N forces, $\Braket{p'q'\alpha'|\mathcal{A}_{123}
V_\text{3N}|pq\alpha}$, are given in a $Jj$-coupled 3N plane-wave
basis of the form
\begin{equation}
\Ket{pq\alpha} = \Ket{pq; \left[(LS)J \left(l\frac{1}{2}\right)j \right] \mathcal{J} \left(T\frac{1}{2}\right)\mathcal{T}}\, ,
\end{equation}
where the relative orbital angular momentum, spin, total angular
momentum, and isospin of particles $1$ and $2$ are labeled by $L$,
$S$, $J$, and $T$ (with $T=1$ in the case of neutron matter). The
quantum numbers $l$ and $j$, respectively, denote the orbital angular
momentum and total angular momentum of particle $3$ relative to the
center of mass of the pair with relative momentum $p$. The quantum
numbers $\mathcal{J}$ and $\mathcal{T}$ are the total 3N angular
momentum and isospin (with $\mathcal{T}=3/2$ here). These 3N matrix
elements are currently available up to N$^3$LO~\cite{Hebe15N3LOpw},
with a large enough truncation on the total three- and two-body total
angular momenta $\mathcal{J} \leqslant 9/2$ and $J \leqslant 6$,
respectively, to obtain well converged 3N Hartree-Fock energies in
neutron and symmetric nuclear
matter~\cite{Hebe15N3LOpw,Dris15asym}. We refer to these references
also for detailed discussions of normal ordering in the partial-wave
basis. The effective NN potential~\eqref{eq:Veff_formal} depends in
general on the total momentum $\vec{P}$ of the two remaining particles
in contrast to a Galilean-invariant NN interaction. At the BCS level,
the paired particles are in back-to-back kinematics and we therefore
have $\vec{P}=0$.

The normal-ordered two-body part of 3N forces can then be combined
with NN interactions:
\begin{equation} \label{eq:NO_Pot}
V_\text{NN+3N}^{\text{as}} = V_\text{NN}^{\text{as}} + \zeta \, \overline{V}_\text{3N}^{\text{as}} \, ,
\end{equation}
where $\zeta$ is a combinatorial factor that depends on the type of
quantity of interest (see Ref.~\cite{Hebe10nmatt} for details). For
the gap equation~\eqref{eq:gap_pw} we find $\zeta=1$ (see
Appendix~\ref{app:sym_fac} for details).

The energy denominator of the gap equation~\eqref{eq:gap_pw} depends
on the single-particle energy $\varepsilon(k)$. We take into account
self-energy corrections to the kinetic energy due to the
interaction~\eqref{eq:NO_Pot}.  In the Hartree-Fock approximation the
single-particle energy is given by
\begin{equation}\label{eq:SPE}
\varepsilon(k) = \frac{k^2}{2m} + \Sigma^{(1)}(k) \, ,
\end{equation}
where $\Sigma^{(1)}(k)$ denotes the spin-averaged Hartree-Fock self-energy~\cite{Hebe10nmatt},
\begin{equation} \label{eq:sigma_HF}
\begin{split}
\Sigma^{(1)}(k_1)&= \frac{1}{2\pi} \int dk_2 \, k_2^2  \int d\cos \theta_{\vec{k}_1,\vec{k}_2} \, n_{\vec{k}_2} \sum_{l,S,J} (2J+1) \\ 
&\quad \times  \Braket{k_{12}/2|V_{llS}^{J}|k_{12}/2} \left(1 - (-1)^{l+S+1} \right) \, ,
\end{split}
\end{equation}
with $k_{12} = |\vec{k}_1-\vec{k}_2|$. For the combinatorial factor
$\zeta$ in the interaction matrix element
$V=V_\text{NN+3N}^{\text{as}}$ in Eq.~\eqref{eq:sigma_HF} we obtain
$\zeta=1/2$ (see Appendix~\ref{app:sym_fac} or
Refs.~\cite{Hebe10nmatt,Srin16PairPNM,Dris15asym}).  The corresponding
effective mass~$m^*$ at the Fermi surface is then given by
\begin{equation} \label{eq:eff_mass}
\frac{m^*(\kF)}{m} = \left( \frac{m}{k} \frac{d \varepsilon(k)}{dk}\right)^{-1}\bigg|_{k=\kF}  \, .
\end{equation}
Our calculations with a free and a Hartree-Fock spectrum serve as a
simple measure for the dependence of $\Delta(\kF)$ on the
single-particle energy.

\subsection{Theoretical uncertainties}
\label{subsec:uncertainty}

An improved approach for estimating theoretical uncertainties based on
the chiral expansion has been proposed in
Refs.~\cite{Epel15improved,Epel15NNn4lo} and applied to few-body
calculations~\cite{Bind16SoANN,Lynn16QMC3N}. In contrast to previous
uncertainty estimates, which involved cutoff variations of nuclear
interactions at a given chiral order, these are based on results at
different chiral orders for a fixed cutoff value. This allows to study
the order-by-order convergence in the chiral expansion for an
observable at a given momentum scale. Currently,
local~\cite{Geze13QMCchi,Geze14long} and
semilocal~\cite{Epel15improved,Epel15NNn4lo} NN potentials are
available up to N$^2$LO and N$^4$LO, respectively, with cutoffs of
$R_0 = (0.8-1.2)\fm$. Semilocal means in this context that only the
long-range part is regularized locally in coordinate space whereas the
short-range part is regularized nonlocally in momentum space.

The contributions to the gap from interaction terms at chiral order $i =
0,2,3,\ldots$ are given by
\begin{equation}
\mathrm{d}\Delta^{(i)} = 
\begin{cases}
\Delta^{(2)} - \Delta^{(0)} & i = 2 \, ,\\
\Delta^{(i)} - \Delta^{(i-1)} & i\geqslant 3 \, ,
\end{cases}
\end{equation}
and are expected to scale like $\bigl(Q(\kF)\bigr)^i$ where
\begin{equation} \label{eq:Q_def}
Q(\kF) = \max \left( \frac{p}{\Lambda_b}, \frac{m_\pi}{\Lambda_b} \right)  \,
\end{equation}
is the ratio of a typical momentum scale $p$ or $m_\pi$ of the system
and the breakdown scale $\Lambda_b$. Since the pairing gap results
from attractive interactions of two particles on the Fermi surface we
use in the following $p=\kF$ for the relative momentum in
Eq.~\eqref{eq:Q_def}. Note that this scaling is in general only
expected to be valid for complete calculations involving all many-body
forces at a given chiral order. In Sec.~\ref{sec:results} we present
results based on local and semilocal interactions without inclusion of
many-body forces. Complete calculations with full uncertainty
estimates will be possible as soon as partial-wave matrix elements of
the corresponding 3N forces are available. For the local and semilocal
NN interactions the breakdown scale was chosen as follows for the
different cutoffs $R_0$ \cite{Epel15improved}:
\begin{equation} \label{eq:L_b_def}
\Lambda_b = 
\begin{cases}
600 \MeV & \text{for}\;R_0 = 0.8,0.9,1.0\fm \, ,\\
500 \MeV & \text{for}\;R_0 = 1.1\fm \, ,\;\text{and}\\
400 \MeV & \text{for}\;R_0 = 1.2\fm  \, .
\end{cases}
\end{equation}

The chiral expansion can be used to define the theoretical
uncertainty~\cite{Epel15NNn4lo,Epel15improved}, where we focus on
uncertainties at N$^2$LO and higher ($i \geqslant 3$),
\begin{equation} \label{eq:our_uncert}
\delta \Delta^{(i)} = \max \limits_{3\leqslant j\leqslant i} \left( Q^{i+1-j} \left|\mathrm{d} \Delta^{(j)}\right|\right) \, .
\end{equation}
We do not show uncertainties at LO and NLO, because at these orders
the scattering phase shifts are not well reproduced at the relevant
momenta, particularly not in the coupled \threePFtwo channel. Note
that, in contrast to Refs.~\cite{Epel15NNn4lo, Epel15improved}, for
the above reason we neglect the LO contributions to the higher-order
uncertainties, and moreover we do not consider a term that ensures
that the next order always lies within the uncertainty band of the
previous order by taking into account information of higher-order
results in the chiral expansion.

As mentioned in Sec.~\ref{subsec:NO_3N_SPE}, the normal ordering is
currently based on 3N forces with nonlocal regulators. Once available,
it will be straightforward to incorporate also local or semilocal 3N
interactions.  Work in this direction is currently in
progress. Following the paradigm to regularize NN and many-body forces
consistently, we do not show results based on local or semilocal NN
forces combined with nonlocal 3N interactions. Instead, we use the
nonlocal N$^3$LO NN potentials EM~500~MeV~\cite{Ente03EMN3LO,
Mach11PR}, EGM~450/500~MeV and EGM~450/700~MeV~\cite{Epel05EGMN3LO}
with the 3N uncertainty estimate governed by variation of the 3N
parameters $c_1,c_3$ and $\Lambda_\text{3N}=(2.0-2.5)\fmi$. As
recommended in Ref.~\cite{Kreb123Nlong}, we take for calculations with
N$^2$LO 3N forces the ranges $c_1=-(0.37-0.73)\GeVi$,
$c_3=-(2.71-3.38)\GeVi$ and with N$^3$LO 3N forces
$c_1=-(0.75-1.13)\GeVi$, $c_3=-(4.77-5.51)\GeVi$. The N$^3$LO 3N
contributions shift $c_1,c_3$ and depend additionally on the LO NN
low-energy constants which we consider consistently with the NN
potentials. A compilation of the values can be found in Table~I of
Ref.~\cite{Krue13N3LOlong}.

\section{Results}
\label{sec:results}

\subsection{Local and semilocal NN potentials}

\begin{figure*}[p]
\includegraphics[page=1,scale=0.95,clip]{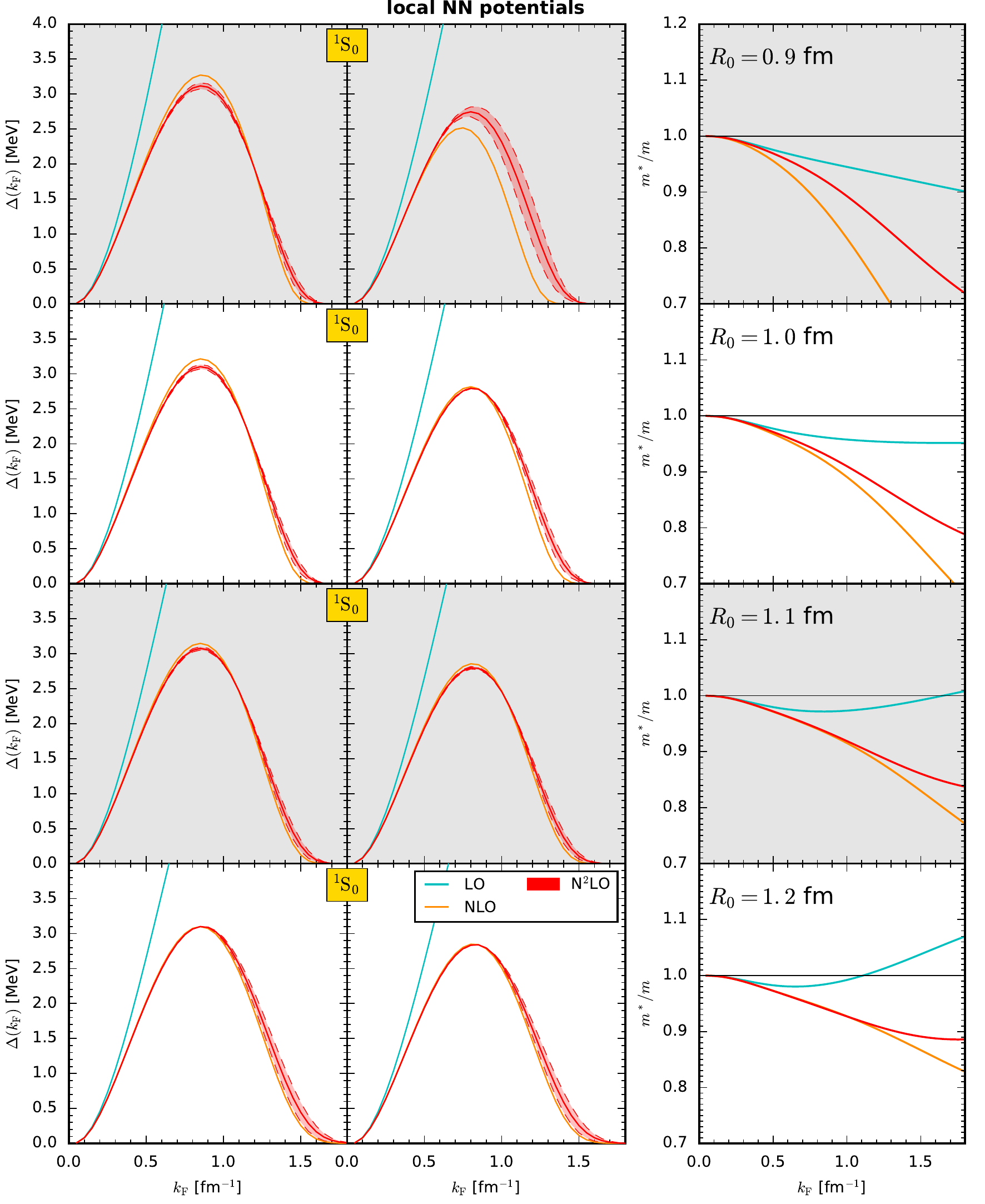}
\caption{\label{fig:1S0_uncert_local}(Color online)
Gap $\Delta$ as a function of Fermi momentum $k_{\rm F}$ in the \oneSzero 
channel for the four local NN potentials with
$R_0=(0.9-1.2)\fm$~(rows) each up to N$^2$LO with a free (left) and a
Hartree-Fock spectrum~(center column), respectively. The third column
shows the effective mass at the Fermi surface corresponding to the
Hartree-Fock spectrum (second column). As discussed in the text, the
uncertainty bands (if present) are given by the color-filled region
between the dashed lines while the actual calculation is depicted by
the solid line. There are no uncertainties shown for LO and NLO; for 
details see text.}
\end{figure*}

\begin{figure*}[p]
\includegraphics[page=1,scale=0.95,clip]{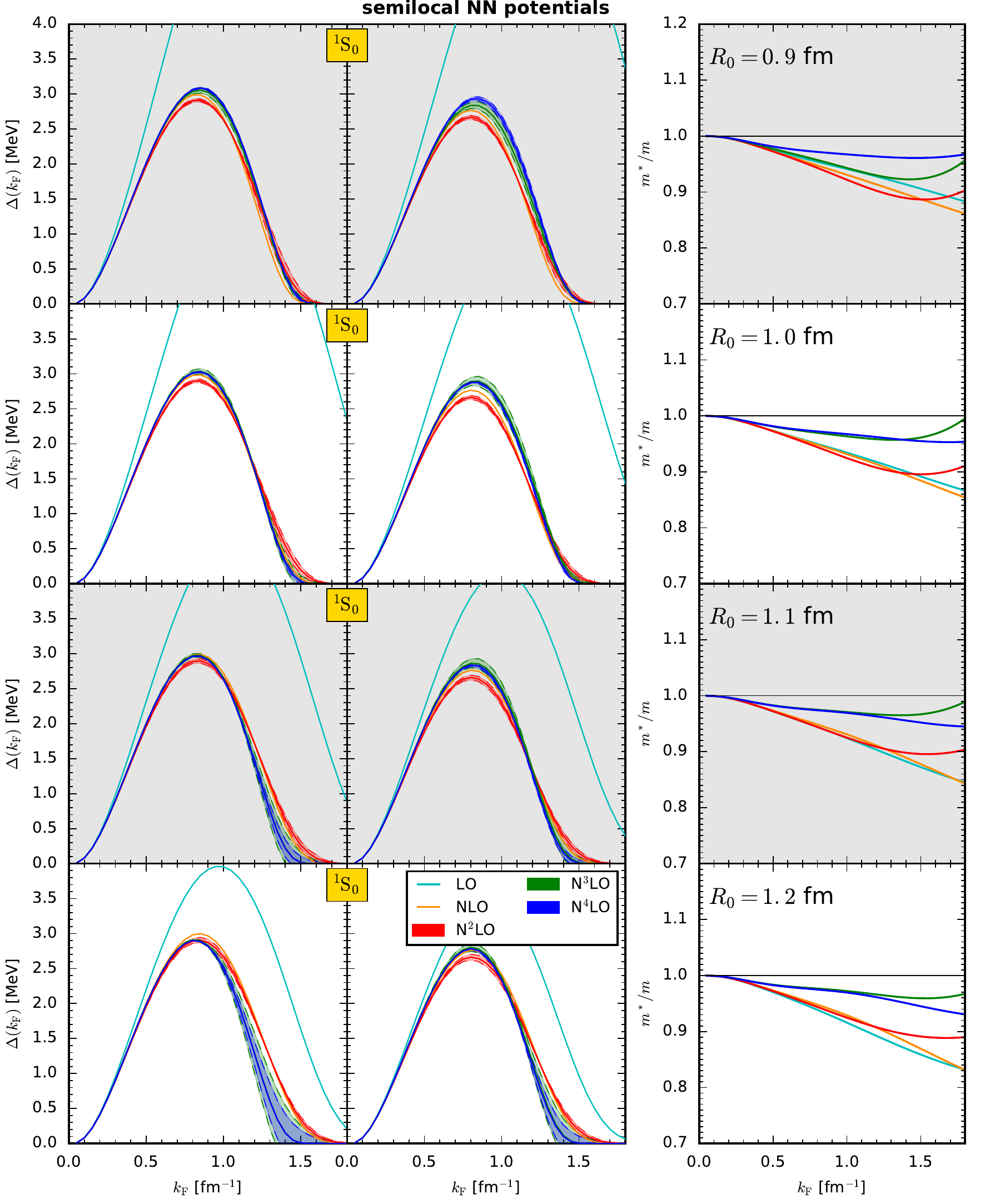}
\caption{\label{fig:1S0_uncert_semilocal}(Color online)
Gap $\Delta$ as a function of Fermi momentum $k_{\rm F}$ in the \oneSzero
channel for the four semilocal NN potentials with
$R_0=(0.9-1.2)\fm$~(rows) each up to N$^4$LO with a free (left) and a
Hartree-Fock spectrum~(center column), respectively. The third column
shows the effective mass at the Fermi surface corresponding to the
Hartree-Fock spectrum (second column). There are no uncertainties shown 
for LO and NLO; for details see text.}
\end{figure*}

We present in Figs.~\ref{fig:1S0_uncert_local} and
\ref{fig:1S0_uncert_semilocal} the gap in the ${}^1S_0$ channel based
on the local and semilocal NN potentials up to N$^2$LO and N$^4$LO,
respectively. Each row corresponds to the regulators $R_0=0.9, 1.0,
1.1$ and $1.2 \fm$ as annotated.  The left (center) column shows the
gap using a free (Hartree-Fock) spectrum. The effective mass
$m^*(\kF)/m$ from the Hartree-Fock spectrum are depicted in the right
column. As discussed in Sec.~\ref{subsec:uncertainty} we assign
uncertainty estimates to the results beyond NLO according to
Eq.~\eqref{eq:our_uncert}. In Figs.~\ref{fig:1S0_uncert_local} and
\ref{fig:1S0_uncert_semilocal} the results for $\Delta(\kF)$ at
different orders are depicted by solid lines, and the uncertainty
bands $\Delta(\kF) \pm \delta \Delta$ are shown as shaded bands whose
boundaries are highlighted by dashed lines. We restrict the bands to
the region of positive energies.

At NLO and beyond we observe that the \oneSzero gap agrees up to
$\kF\sim(0.6-0.8)\fmi$, depending only slightly on the regulator for
local potentials. As investigated in detail, e.g., in
Ref.~\cite{Hebe07Pairing}, the pairing gaps are strongly constrained
by phase shifts. The LO gaps are therefore expected to be
different. For $R_0 \geqslant 1.0\fm$ we find that the gaps at N$^3$LO
and N$^4$LO agree well over the entire density range.  Generally, the
gap uncertainties based on Eq.~(\ref{eq:our_uncert}) are very small
for the highest chiral orders. However, we emphasize that the gap
uncertainties only include contributions from the chiral expansion,
whereas neglected higher-order many-body corrections are not
assessed.

In addition, we find that the sensitivity of the pairing gap to the
energy spectrum is rather small and affects mainly the maximum value
of the gap. For both, local and semilocal potentials we find
$\Delta_\text{max} \sim (2.7-3.1)\MeV$ at $\kF\sim(0.8-0.9)\fmi$ for
the highest chiral order and all cutoffs. The rather small suppression
due to the spectrum can be directly understood based on the fact that
the ratio $m^*(\kF)/m$ is close to one for all regulators and chiral
orders (right columns).

\begin{figure*}[p]
\includegraphics[page=1,scale=0.95,clip]{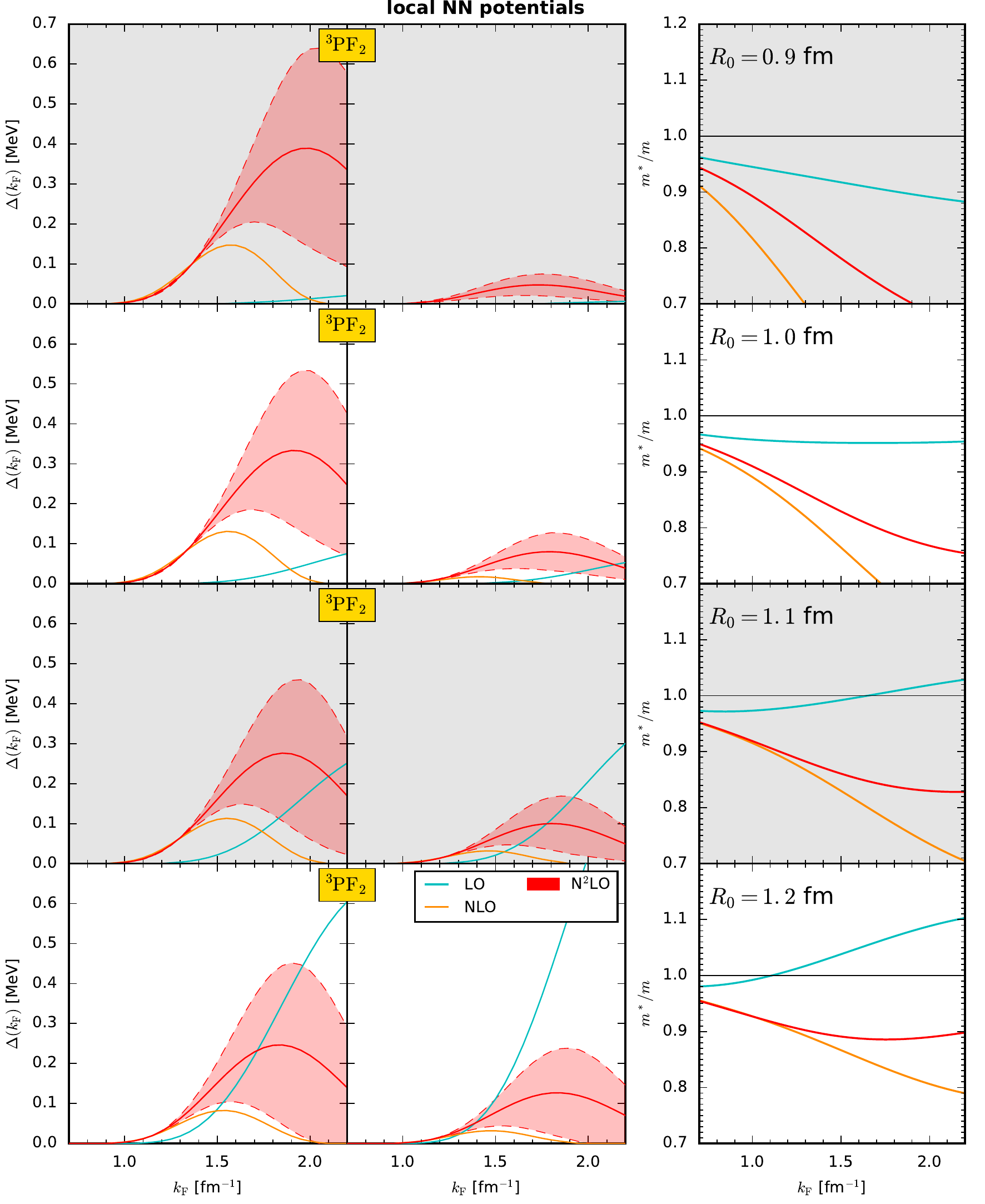}
\caption{\label{fig:3PF2_uncert_local}(Color online)
Gap $\Delta$ as a function of Fermi momentum $k_{\rm F}$ in the
\threePFtwo channel for the four local NN potentials with
$R_0=(0.9-1.2)\fm$~(rows), each up to N$^2$LO with a free (left) and a
Hartree-Fock spectrum~(center column), respectively. The third row
shows the effective mass at the Fermi surface corresponding to the
Hartree-Fock spectrum (second column). As discussed in the text, the
uncertainty bands (if present) are given by the color-filled region
between the dashed lines while the actual calculation is depicted by
the solid line. There are no uncertainties shown for LO and NLO; for
details see text.}
\end{figure*}

\begin{figure*}[p]
\includegraphics[page=1,scale=0.95,clip]{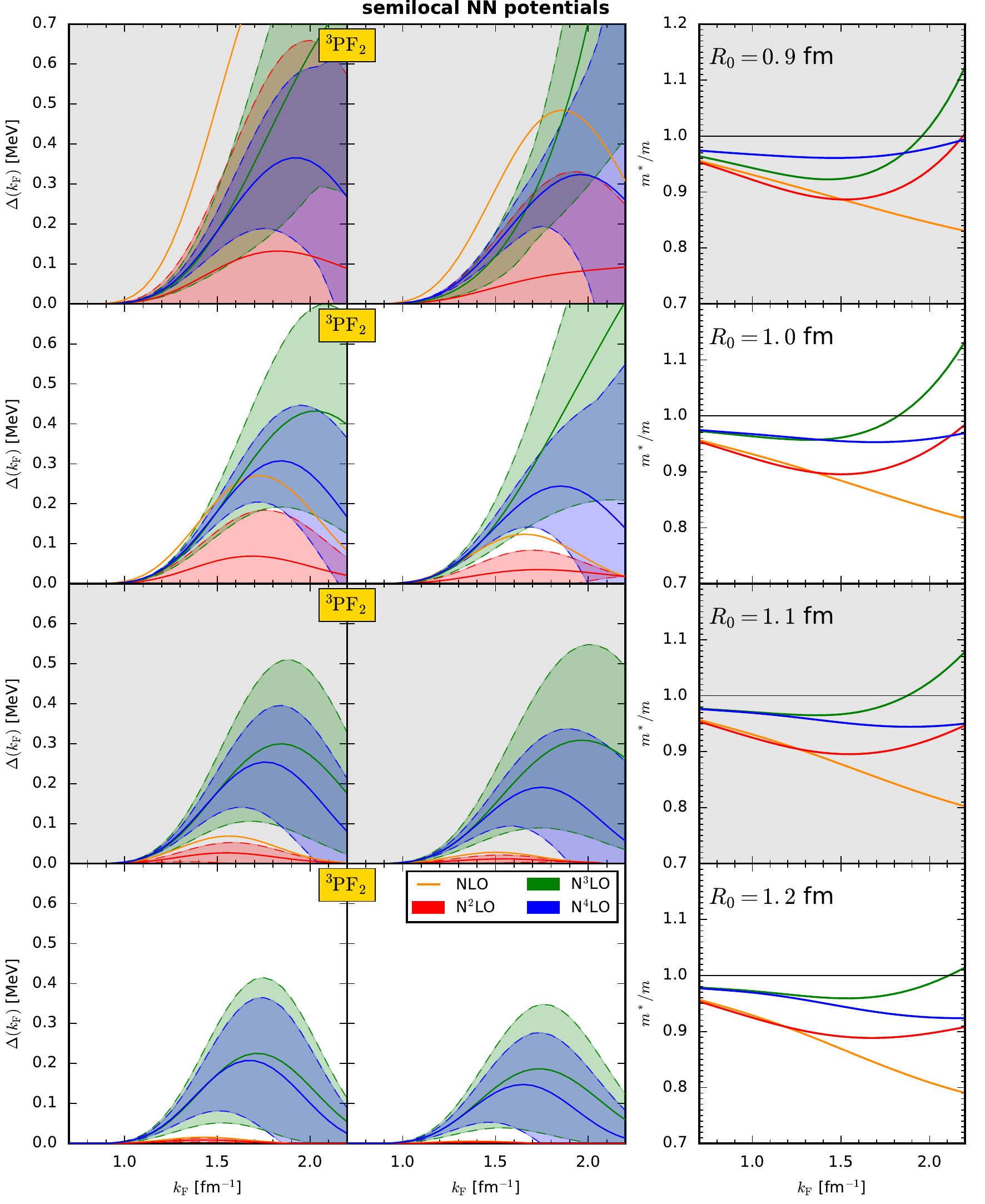}
\caption{\label{fig:3PF2_uncert_semilocal}(Color online)
Gap $\Delta$ as a function of Fermi momentum $k_{\rm F}$ in the 
\threePFtwo channel for the four semilocal NN potentials with
$R_0=(0.9-1.2)\fm$~(rows), each up to N$^4$LO with a free (left) and
a Hartree-Fock spectrum~(center column), respectively. The third row
shows the effective mass at the Fermi surface corresponding to the
Hartree-Fock spectrum (second column). As discussed in the text, the
uncertainty bands are given by the color-filled region between the
dashed lines while the actual calculation is depicted by the solid
line. There are no uncertainties shown for LO and NLO; for
details see text.}
\end{figure*}

In Figs.~\ref{fig:3PF2_uncert_local} and
\ref{fig:3PF2_uncert_semilocal} we show the \threePFtwo gap based on
the same NN potentials. Since \threePFtwo pairing takes place at
larger densities than in the \oneSzero channel, the uncertainties are
much larger. The maximum of the LO pairing gap for the local
potentials changes significantly with increasing $R_0$, indicating
that the results are strongly affected by regulator artifacts at this
order. On the other hand, the pairing gap for the semilocal potentials
at LO is vanishing for all densities and cutoff values and therefore
not shown in Fig.~\ref{fig:3PF2_uncert_semilocal}. These results
reflect the poor description of the phase shifts at this order, from
only the one-pion-exchange interaction at this order for the semilocal
case.

At higher chiral orders it is not straightforward to extract robust
quantitative trends for the \threePFtwo gap. In general, the gap opens
around densities of $\kF \sim 1 \fmi$ for all considered interactions.
For the semilocal potentials the results at N$^3$LO and N$^4$LO agree
well up to $\kF \sim 1.6\fmi$. Also the corresponding uncertainty
bands strongly overlap in this density region. We find the maximum gap
values at N$^2$LO and higher orders in the density range $\kF=
(1.6-2.1)\fmi$ for all interactions.  Overall, the large uncertainties
at high densities reflect the regulator dependences and the breakdown
of the chiral expansion. In particular, for a Fermi momentum $\kF =
2.0 \fmi$ the expansion parameter $Q(\kF)$ of Eq.~\eqref{eq:Q_def}
is
\begin{equation}
Q(2.0 \fmi) =
\begin{cases}
0.66 & \text{for}\;R_0 = 0.8,0.9,1.0\fm \, ,\\
0.79 & \text{for}\;R_0 = 1.1\fm \, ,\;\text{and}\\
0.99 & \text{for}\;R_0 = 1.2\fm  \, .
\end{cases}
\end{equation}
Clearly, it is not obvious that the chiral expansion is efficient
anymore in this density regime.

\subsection{N$^2$LO and N$^3$LO 3N forces}

\begin{figure*}[t]
\includegraphics[page=1,width=\textwidth,clip]{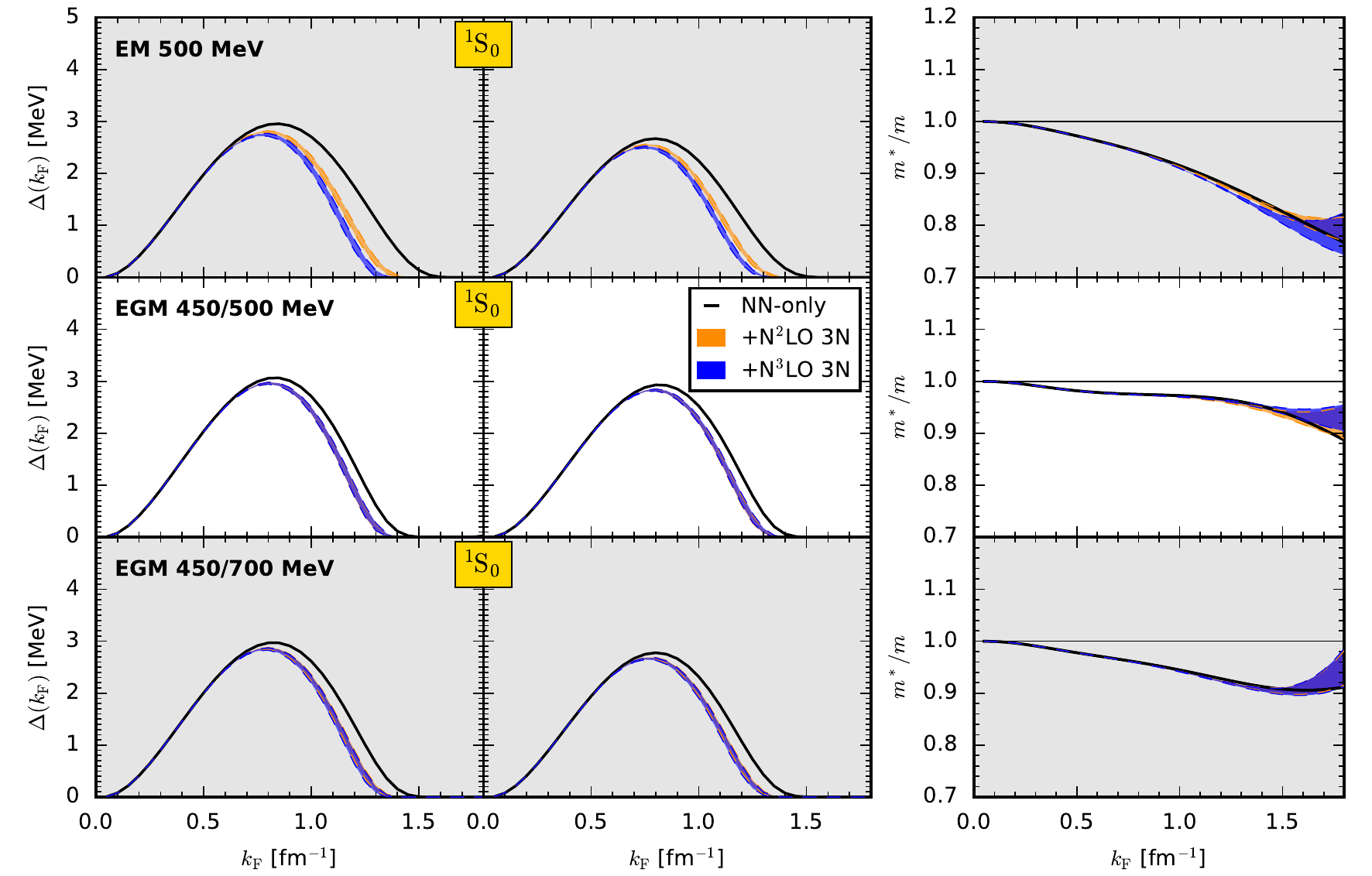}
\caption{\label{fig:1S0_3N}(Color online)
Gap $\Delta$ as a function of Fermi momentum $k_{\rm F}$ in the
\oneSzero channel with a free (left) and a Hartree-Fock spectrum
(center column) for the N$^3$LO NN potentials EM~500~MeV (first),
EGM~450/500~MeV (second) and EGM~450/700~MeV (third row).  The third
column depicts the effective mass at the Fermi surface corresponding
to the Hartree-Fock spectrum.  The NN-only results are shown by the
black-solid lines. The uncertainty bands for N$^2$LO and N$^3$LO are
determined by variations of the 3N parameters $c_1, c_3$ and
$\Lambda_\text{3N}$ as discussed in the text.}
\end{figure*}

\begin{figure*}[t]
\includegraphics[page=1,width=\textwidth,clip]{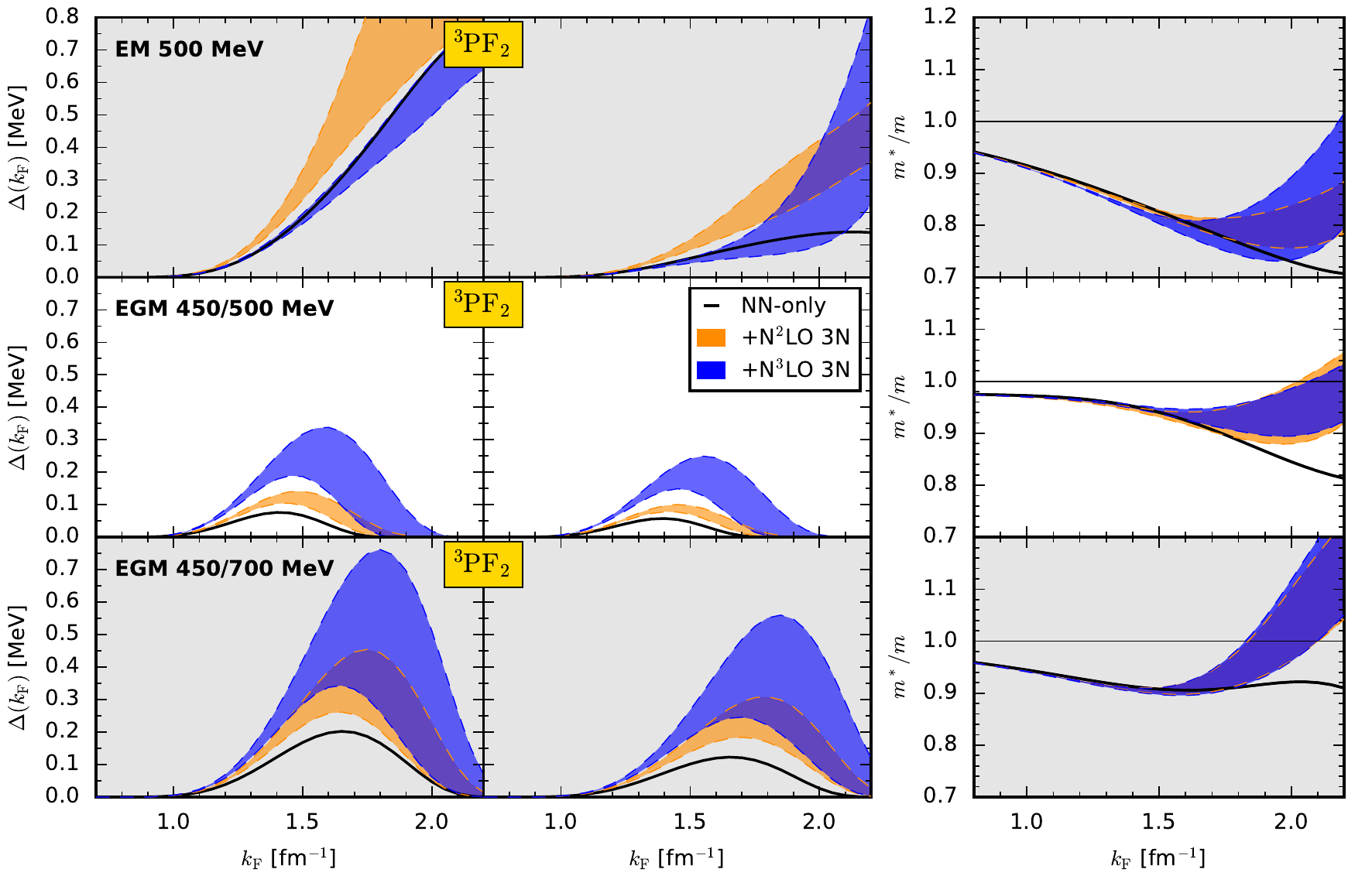}
\caption{\label{fig:3PF2_3N}(Color online)
Gap $\Delta$ as a function of Fermi momentum $k_{\rm F}$ in the
\threePFtwo channel with a free (left) and a Hartree-Fock spectrum
(center column) for the N$^3$LO NN potentials EM~500~MeV (first),
EGM~450/500~MeV (second) and EGM~450/700~MeV (third row).  The third
column depicts the effective mass at the Fermi surface corresponding
to the Hartree-Fock spectrum.  The NN-only results are shown by the
black-solid lines. The uncertainty bands for N$^2$LO and N$^3$LO are
determined by variations of the 3N parameters $c_1, c_3$ and
$\Lambda_\text{3N}$ as discussed in the text.}
\end{figure*}

We also study the pairing gaps based on three nonlocal NN potentials
at N$^3$LO combined with contributions from N$^2$LO and N$^3$LO 3N
forces. The results are shown in Figs.~\ref{fig:1S0_3N}
and~\ref{fig:3PF2_3N} in the \oneSzero and \threePFtwo channel,
respectively. The rows correspond to the NN potential
EM~500~MeV~\cite{Ente03EMN3LO}, EGM~450/500~MeV and
EGM~450/700~MeV~\cite{Epel05EGMN3LO}, as annotated. The left and
center columns show energy gaps using a free and Hartree-Fock
spectrum, whereas the right column shows the corresponding
Hartree-Fock effective mass. NN-only results are shown by black solid
lines, with the inclusion of the leading (subleading) 3N forces by
orange (blue) bands. As discussed in Sec.~\ref{subsec:uncertainty},
the uncertainty bands are obtained by variations of the 3N parameters
$c_1,c_3$ and $\Lambda_\text{3N}$.

Figs.~\ref{fig:1S0_uncert_local}, \ref{fig:1S0_uncert_semilocal}
and~\ref{fig:1S0_3N} show that the \oneSzero gaps at N$^3$LO without
3N forces are in good agreement. This observation can be traced back
to the well-reproduced phase shifts at this order. Contributions from
3N forces do not change the results for the pairing gaps at low
densities, $\kF \lesssim (0.7-0.8) \fmi$, and only lead to a minor
suppression at higher densities. The uncertainty bands including 3N
forces are very small for all potentials at N$^2$LO as well as
N$^3$LO. In addition, self-energy contributions to the single-particle
energies are small.

In Fig.~\ref{fig:3PF2_3N} we show the corresponding results for the
\threePFtwo channel. Since the relevant densities are larger than in
the \oneSzero channel, the impact of 3N forces is generally larger for
the pairing gap and also for the effective mass. We observe
nonvanishing gaps for the three investigated NN potentials for all
three cases considered. In contrast to the \oneSzero channel the
inclusion of 3N forces typically provides additional attraction and
hence increases the pairing gap, except for the EM~500~MeV potential
with subleading 3N forces. As shown in the right column, 3N
contributions generally tend to enhance the effective mass (see also
Ref.~\cite{Hebe10nmatt}), even to values larger than one at the
Hartree-Fock level. In general, we find that the results for the
\threePFtwo pairing gaps differ significantly for the various
potentials and that it is delicate to extract robust quantitative
predictions based on our results.

\section{Summary and Outlook}
\label{sec:summary}

In this paper, we have studied solutions of the BCS gap equation in
the \oneSzero and the \threePFtwo channel based on a broad range of
nuclear interactions derived within chiral EFT at different chiral
orders. We benchmarked and optimized two different algorithms that
allow a reliable and accurate solution of the nonlinear BCS
equation. With these advances, we studied the gap based on local NN
interactions~\cite{Geze13QMCchi,Geze14long} up to N$^2$LO and
semilocal NN interactions~\cite{Epel15improved,Epel15NNn4lo} up to
N$^4$LO for the four coordinate-space cutoffs $R_0=0.9, 1.0, 1.1$ and
$1.2$~fm. At the highest chiral orders the results in the \oneSzero
channel agree for all interactions over the entire density region. The
\oneSzero pairing gap reaches a maximum around $\kF=(0.8-0.9)\fmi$
with $\Delta_{\text{max}} = (2.9-3.3)\MeV$ for a free single-particle
spectrum and a suppression of about $\sim 0.3\MeV$ when including
self-energy corrections in the Hartree-Fock approximation.

In the triplet channel \threePFtwo the situation is much less
clear. The gaps generally open at densities of $\kF \sim
(0.9-1.0)\fmi$ for all interactions. Beyond this density the results
depend on details of the interactions and the chiral order. At the
highest chiral orders we observe a gap maximum at densities in the
region $\kF = (1.7 - 1.9)\fmi$ with $\Delta_{\text{max}} <
0.4\MeV$. However, we emphasize that these Fermi-momentum scales are
already close to the EFT breakdown scale of the corresponding
interactions. Consequently, the observed strong regulator dependence
is not surprising.

For the estimate of theoretical order-by-order uncertainties of the
Hamiltonian we followed the method first presented in
Ref.~\cite{Epel15improved} with two modifications. We obtained very
small uncertainties for the \oneSzero channel for all densities, but
sizable uncertainties in the \threePFtwo channel. In the latter case
the uncertainty bands at successive chiral orders are generally not
entirely overlapping. However, at N$^3$LO and N$^4$LO we find that the
bands are of comparable size and overlapping. We emphasize that our
calculations at N$^2$LO, N$^3$LO and N$^4$LO are not complete since no
3N forces have been taken into account for these interactions. Hence,
the analysis should be revisited as soon as the calculation of the
corresponding local 3N partial-wave matrix elements have been
completed. This is work in progress.

In addition, we also investigated the impact of 3N forces on the
pairing gap for nonlocal N$^3$LO potentials. Taking advantage of
recent developments for including 3N forces in a partial-wave
basis~\cite{Hebe15N3LOpw,Dris15asym}, we were able to incorporate for
the first time subleading 3N contributions in the gap equation via
normal ordering. We found only small repulsive effects from 3N forces
in the singlet channel \oneSzero, whereas in the \threePFtwo channel
the effects from 3N forces are larger and lead to attractive
contributions in most cases. Also for these interactions, we find
significant regulator dependences in the \threePFtwo channel.

We conclude that due to the high densities of the \threePFtwo gaps,
which are reaching the limit of the employed chiral EFT interactions,
it is not possible to draw final quantitative conclusions on the size
of the \threePFtwo gap in neutron matter. However, we have observed
nonvanishing gaps for all employed realistic NN potentials, also when
including 3N contributions. We further emphasize that the
contributions from higher many-body corrections beyond the BCS
approximation have not been taken into account and are known to be
significant~\cite{Geze14pairing,Schw03noncentral}, although their
quantitative assessment is especially challenging in the \threePFtwo
channel.

The methods discussed in this paper can be used for improved studies
of pairing gaps in the future. In particular, the advanced treatment
of 3N forces in terms of partial waves allows to handle in a
straightforward way arbitrary partial-wave decomposed 3N forces. In
addition, it is also possible to perform calculations based on
consistently-evolved NN and 3N forces~\cite{Hebe12msSRG} via the
similarity renormalization group (SRG).  This is in particular of
interest when taking into account many-body corrections beyond the BCS
approximation in calculations of the pairing gap since SRG-evolved
forces are expected to exhibit an improved many-body convergence.

\begin{acknowledgments}

We thank E.~Epelbaum, J.~W.~Holt, S.~Ramanan, V.~Som\`a, and
S.~Srinivas for fruitful discussions. This work was supported in part
by the European Research Council Grant No.~307986 STRONGINT and the
Deutsche Forschungsgemeinschaft through Grant SFB~1245.

\end{acknowledgments}

\appendix

\section{Partial-wave decomposition}
\label{app:PW}

In this appendix we briefly review the partial-wave decomposition of
the gap equation~\eqref{eq:gap_pw} and specify the conventions used
in this work.  Following Refs.~\citep{Clar03TripPair, Taka70PNM1,
  Khod01PairPNM} we decompose the gap matrix in the form
\begin{equation}
\label{appeq:exp_Gap}
\Delta_{\alpha\alpha'}(\vec{k}) = \sum \limits_{\substack{l,S\\J,M}} \sqrt{\frac{8\pi}{2J+1}} \, \Delta_{lS}^{JM}(k) \left(G_{lS}^{JM}(\hat{\vec{k}})\right)_{\alpha\alpha'} \,,
\end{equation}
and accordingly the nuclear interaction
\begin{equation}
\label{appeq:exp_Pot}
\begin{split}
&(4\pi)^{-2}\braket{\vec{k}\alpha\alpha'|V|\vec{k}' \beta\beta'} \\
&= \sum \limits_{\substack{l,l',S\\J,M}} i^{l'-l} \left(G_{lS}^{JM}(\hat{\vec{k}})\right)_{\alpha\alpha'} \left(G_{l'S}^{JM}(\hat{\vec{k}}')\right)_{\beta\beta' }^* V_{ll'S}^{J(M)}(k,k') \,,
\end{split}
\end{equation}
with
\begin{equation}
\left(G_{lS}^{JM}(\hat{\vec{k}})\right)_{\alpha\alpha'} = \sum \limits_{m,m_S} \clebschG{1/2}{\alpha}{1/2}{\alpha'}{S}{m_S} \clebschG{l}{m}{S}{m_S}{J}{M} Y_l^m(\hat{\vec{k}}) \, .
\end{equation}
These functions obey the orthogonality relations
\begin{equation}
\label{appeq:G_identity}
\begin{split}
\int d\Omega_\vec{k}  &  \sum_{\beta,\beta'}  \left[ \left( G_{l'S}^{JM}(\hat{\vec{k}}')\right)_{\beta\beta'}^* \left(G_{l''S'}^{J'M'}(\hat{\vec{k}}')\right)_{\beta\beta'} \right] \\
&= \delta_{l l'} \delta_{M M'} \delta_{J J'} \delta_{S S'} \, .
\end{split}
\end{equation}
The $J$-dependent factor in Eq.~\eqref{appeq:exp_Gap} is chosen such
that the gap equation in partial-wave representation takes a
particularly simple form.  Inserting Eqs.~\eqref{appeq:exp_Gap}
and~\eqref{appeq:exp_Pot} in the gap equation~\eqref{eq:gap_start}
leads to
\begin{equation}
\label{appeq:Gab_ang}
\begin{split}
&(4\pi)^{-2}\sum \limits_{\substack{l,S\\J,M}}  \frac{\Delta_{lS}^{JM}(k)}{\sqrt{2J+1}} \left(G_{lS}^{JM}(\hat{\vec{k}})\right)_{\alpha\alpha'} \\
&= -  \int \frac{dk' \, k'^2}{(2\pi)^3}   \sum \limits_{\substack{l,l',J,M,S \\ l'',J',M',S'}} i^{l'-l} \left(G_{lS}^{JM}(\hat{\vec{k}})\right)_{\alpha\alpha'} \\
&\quad\times V_{ll'S}^{J(M)}(k,k') \frac{\Delta_{l''S'}^{J'M'}(k')}{\sqrt{2J'+1}}  \\
&\quad \times \int d\Omega_{\vec{k}'}   \frac{ \sum_{\beta,\beta'} \left[ \left( G_{l'S}^{JM}(\hat{\vec{k}}')\right)_{\beta\beta'}^* \left(G_{l''S'}^{J'M'}(\hat{\vec{k}}')\right)_{\beta\beta'}\right]}{2 \sqrt{\xi^2(k')+\frac{1}{2} \text{Tr}  \left[ \Delta \Delta^\dagger\right](\vec{k}') }} \, .
\end{split}
\end{equation}
This equation can be simplified significantly by averaging the energy
gap in the denominator over all angles, specifically
\begin{equation}
\label{appeq:angle_av}
\begin{split}
\frac{1}{2}\text{Tr} \left[ \Delta \Delta^\dagger \right] &\xrightarrow{\text{av.}} \frac{1}{2} \int \frac{d\Omega_\vec{k}}{4\pi} \text{Tr} \left[ \Delta \Delta^\dagger \right] \\
&= \sum \limits_{l,S,J} |\Delta_{lS}^{J}(k')|^2 \equiv D^2(k) \, .
\end{split}
\end{equation}
Here we summed over all $M$ states and used
identity~\eqref{appeq:G_identity}. Projecting out the components in
Eq.~\eqref{appeq:Gab_ang} leads to the partial-wave decomposed gap
equation
\begin{equation}
\label{appeq:gap_pw}
\Delta_{lS}^{J}(k) = - \int_{0}^{\infty}  \frac{dk' \, k'^2}{\pi}  \sum \limits_{l'}  \frac{ i^{l'-l} V_{ll'S}^{J}(k,k') \Delta_{l'S}^{J}(k')}{\sqrt{\xi^2(k')+  \sum \limits_{\tilde{l},\tilde{S},\tilde{J}} |\Delta_{\tilde{l}\tilde{S}}^{\tilde{J}}(k')|^2 }} \, .
\end{equation}

\section{Normal-ordering symmetry factors} \label{app:sym_fac}

In this section we discuss the symmetry factor $\zeta$ that appears in
the interaction kernel in Eq.~(\ref{eq:NO_Pot}) for normal-ordered 3N
contributions in the normal self-energy $\Sigma$ and the anomalous
self-energy $\Delta$.  For this we consider a general Hamiltonian of
the form
\begin{equation}
\hat{H} = \hat{T} + \hat{V}_{\text{NN}} + \hat{V}_{\text{3N}} \, ,
\end{equation}
where $\hat{T}$ represents the kinetic energy, $\hat{V}_{\text{NN}}$
all two-nucleon interactions and $\hat{V}_{\text{3N}}$ three-nucleon
interactions. By using Wick's theorem we can recast the Hamiltonian
exactly in an equivalent form by normal ordering all operators with
respect to a given reference state. For the treatment of superfluid
systems it is convenient to choose the BCS state as reference
state. We represent $\hat{V}_{\text{NN}}$ and $\hat{V}_{\text{3N}}$ in
terms of antisymmetrized matrix elements:
\begin{align}
\hat{V}_{\text{NN}} &= \frac{1}{4} \sum_{ijkl} \left< i j | V^{\rm as}_{\text{NN}} | k l \right> \hat{a}_i^{\dagger} \hat{a}_j^{\dagger} \hat{a}_l \hat{a}_k \, , \\
\hat{V}_{\text{3N}} &= \frac{1}{36} \sum_{ijklmn} \left< i j k | V^{\rm as}_{\text{3N}} | l m n \right> 
\hat{a}_i^{\dagger} \hat{a}_j^{\dagger} \hat{a}_k^{\dagger} \hat{a}_n \hat{a}_m \hat{a}_l \, ,
\end{align} 
where the indices represent generic single-particle quantum numbers.
When applying Wick's theorem with respect to a BCS reference state it
is important to note that both normal contractions (connecting a
creation operator with an annihilation operator) as well as anomalous
contractions (connecting two creation or two annihilation operators)
contribute. For the normal self-energy $\Sigma$ the relevant
contractions are of the form
\begin{align}
&\frac{1}{4} \sum_{ijkl} \left< i j | V^{\rm as}_{\text{NN}} | k l \right> 
\contraction[1.0ex]{\hat{a}_i^{\dagger}}{\hat{a}_j^{\dagger}}{}{\hat{a}_l}
\hat{a}_i^{\dagger} \hat{a}_j^{\dagger} \hat{a}_l \hat{a}_k \label{eq:contractions_Sigma_NN} \, , \\
&\frac{1}{36} \sum_{ijklmn} \left< i j k | V^{\rm as}_{\text{3N}} | l m n \right> 
\contraction[1.0ex]{\hat{a}_i^{\dagger} \hat{a}_j^{\dagger}}{\hat{a}_k^{\dagger}}{}{\hat{a}_n}  
\contraction[1.5ex]{\hat{a}_i^{\dagger}}{\hat{a}_j^{\dagger}}{\hat{a}_k^{\dagger} \hat{a}_n}{\hat{a}_m}
\hat{a}_i^{\dagger} \hat{a}_j^{\dagger} \hat{a}_k^{\dagger} \hat{a}_n \hat{a}_m \hat{a}_l \, , \label{eq:contractions_Sigma_3N} 
\end{align} 
whereas for the anomalous self-energy $\Delta$ the relevant
contractions take the form
\begin{align}
&\frac{1}{4} \sum_{ijkl} \left< i j | V^{\rm as}_{\text{NN}} | k l \right> 
\contraction[1.0ex]{}{\hat{a}_i^{\dagger}}{}{\hat{a}_j^{\dagger}}
\hat{a}_i^{\dagger} \hat{a}_j^{\dagger} \hat{a}_l \hat{a}_k \label{eq:contractions_Delta_NN} \,,  \\
&\frac{1}{36} \sum_{ijklmn} \left< i j k | V^{\rm as}_{\text{3N}} | l m n \right> 
\contraction[1.0ex]{\hat{a}_i^{\dagger} \hat{a}_j^{\dagger}}{\hat{a}_k^{\dagger}}{}{\hat{a}_n}  
\contraction[1.0ex]{}{\hat{a}_i^{\dagger}}{}{\hat{a}_j^{\dagger}}{}
\hat{a}_i^{\dagger} \hat{a}_j^{\dagger} \hat{a}_k^{\dagger} \hat{a}_n \hat{a}_m \hat{a}_l \, .\label{eq:contractions_Delta_3N} 
\end{align}
Since the interaction operators are represented in terms of
antisymmetrized matrix elements all different possible choices of
picking creation or annihilation operators are equivalent and just
lead to combinatoric factors.  Hence, in order to determine $\zeta$ it
is necessary to determine the number of different contractions $c_N$
for Eqs.~(\ref{eq:contractions_Sigma_NN}) to
(\ref{eq:contractions_Delta_3N}). We obtain: $c_N = 4$ for
(\ref{eq:contractions_Sigma_NN}), $c_N = 18$ for
(\ref{eq:contractions_Sigma_3N}), $c_N = 1$ for
(\ref{eq:contractions_Delta_NN}) and $c_N = 9$ for
(\ref{eq:contractions_Delta_3N}).  Combining these combinatoric
factors with the prefactors $1/4$ and $1/36$ of the NN and 3N
interactions we directly obtain $\zeta=1/2$ for $\Sigma$ and $\zeta=1$
for $\Delta$. We also note that in the present work we approximate the
normal contractions in (\ref{eq:contractions_Delta_3N}) by their
contributions in normal systems. It has been shown in
Ref.~\cite{Carb14SCGFdd} that the inclusion of correlations in the
reference state has only very small effects on the matrix elements of
the normal-ordered 3N contributions for nuclear matter
calculations. In addition to contributions from normal contractions in
(\ref{eq:contractions_Delta_3N}) we also obtain nonvanishing
contributions from multiple anomalous contractions. However, these
contributions are small since such terms only include contributions
from momenta around the Fermi surface and are of higher order in the
gap.

\clearpage
\bibliographystyle{apsrev4-1}
\bibliography{../../../strongint/strongint}

\end{document}